\documentclass[a4paper,twocolumn]{revtex4}
\usepackage{graphicx}% Include figure files
\usepackage{amssymb}%
\usepackage{amsmath}
\usepackage{epsfig}
\usepackage{wasysym}
\usepackage{ulem}
\usepackage{color}
\usepackage{verbatim}
\usepackage[utf8]{inputenc}
\usepackage{hyperref}

\begin{document}
\title{Aubry transition with small distortions}
\author{O. C\'epas and P. Qu\'emerais}
\affiliation{Institut N\'eel, CNRS, Universit\'e Grenoble Alpes, Grenoble, France }
\date{\today}

\begin{abstract}

  We show that when the Aubry transition occurs in incommensurately
  distorted structures, the amplitude of the distortions is not
  necessarily large as suggested by the standard Frenkel-Kontorova
  mechanical model. By modifying the shape of the potential in such a
  way that the mechanical force is locally stronger (\textit{i.e.}
  increasing the nonlinearities), the transition may occur at a small
  amplitude of the potential with small distortions.  A ``phason'' gap
  then opens, while the phonon spectrum resembles a standard
  undistorted spectrum at higher energies. This may explain the
  existence of pinned phases with very small distortions as experimentally
  observed in charge-density waves.
\end{abstract}
\pacs{PACS numbers:}

\maketitle
\section{Introduction}

The Aubry transition is an equilibrium phase transition at zero
temperature between two distinct incommensurately modulated phases,
driven by changes in model couplings~\cite{Aubry,aubry_ledaeron}. The
two phases differ by their degeneracies. One phase, called sliding
phase, has a continuous degenerate manifold of ground states which
allows their sliding at no energy cost. In the other phase, called
pinned phase, the degeneracy is lifted and the ground state manifold
becomes discontinuous (and is the Cantor space of
Aubry-Mather~\cite{aubry_ledaeron, mather}). In this case, the
symmetry-related ground states are separated by energy barriers, which
make them pinned in space. Aubry transition can thus be viewed as a
pinning transition.  It was originally called the transition by
breaking of analyticity~\cite{Aubry,aubry_ledaeron} because the
envelope function of the modulation (also called the hull function)
changes from continuous in the sliding phase to discontinuous in the
pinned phase when nonlinear effects in the model become large
enough. Originally, Aubry transition has been discussed in the
one-dimensional mechanical Frenkel-Kontorova
model~\cite{Aubry,aubry_ledaeron} and later in various models where
two length scales are in competition~\cite{FK,Peyrard}.

The question of its experimental relevance is still largely
open. Given the generality of the model, it has been claimed to apply
in different contexts: it has been invoked in the question of friction
and the possibility to have ``superlubric'' sliding phases, for
example in two rotated graphene planes that may slide one over the
other with low friction~\cite{Dew}, at incommensurate boundaries of
solids~\cite{lancon2} or for a tip sliding over a surface either in a
stick-slip or continuous manner~\cite{soco}. More recently it has been
observed and discussed in artificial systems of cold atoms subjected
to a periodic optical potential~\cite{Bylinskii}, or in
two-dimensional colloidal monolayers~\cite{Brazda}.  The first
application of Aubry's theory was discussed for incommensurate
charge-density waves in some
solids~\cite{AubryLeDaeron,AubryQuemerais}, which are pinned in the
absence of external electric field~\cite{monceau_book}. It was argued
that the observed pinning may be an intrinsic effect and the
consequence of the Aubry
transition~\cite{AubryLeDaeron,AubryQuemerais}.  However, the
distortions measured experimentally are generally small, of the order
of a few percents of the lattice spacing~\cite{pouget_canadell}. On
the other hand, in the Aubry pinned phase (above the transition at
strong enough coupling to the potential), the distortions are
predicted to be large. For example, in the Frenkel-Kontorova model,
they are typically of the order of tens of percent of the bond length
at the transition and even larger above the
transition~\cite{AubryLeDaeron,AubryQuemerais}.  The same phenomenon
occurs in electronic models of charge-density
waves~\cite{AubryLeDaeron,AubryQuemerais,CQ}. This discrepancy in the
distortions, of an order of magnitude, makes it difficult to reconcile
Aubry's theory with experiments. As a consequence extrinsic sources of
pinning, \textit{i.e.} pinning by impurities, have been invoked to
explain charge-density waves, but they may not be necessary.  The
theoretical issue which we study here by introducing a modified
Frenkel-Kontorova model, is to understand whether pinning is
\textit{necessarily} accompanied by large distortions, or, in other
words, if it is possible to have an incommensurately distorted phase,
with small distortions, yet pinned.

The question of disentangling distortions and pinning is not simple as
both occur as consequences of the presence of the nonlinear potential
in the standard Frenkel-Kontorova model. By introducing a second
length scale, in competition with the first, the potential distorts
the regular structure.  At the same time, by breaking the translation
symmetry that ensures the existence of the sliding phase, the
potential induces pinning.  It is further known that the latter occurs
above a threshold of the amplitude of the nonlinear potential. On the
other hand, the sliding incommensurate phase remains stable below this
threshold as a consequence of Kolmogorov-Arnold-Moser (KAM)
theorem~\cite{Scott} which thus generally prevents a pinned phase with
small distortions.

Let us consider two extreme and opposite situations. First, there are
models, such as that of Brazovskii, Dzyaloshinskii and
Krichever~\cite{BDK}, which exhibit a form of 'super-stability': the
sliding phase is stable for all values of the nonlinear coupling. It
is then impossible to have a pinned phase (\textit{a fortiori} with
small distortions). This behavior is in fact special, it is the
consequence of the integrability of the model: once the integrability
is broken, the Aubry transition occurs~\cite{CQ}. Second, if the
conditions of the KAM theorem are not fulfilled by the nonlinear term,
there is no mathematical ground to have a sliding phase: the threshold
of the Aubry transition could vanish.  A first condition to apply KAM
theorem is that the perturbation of the integrable dynamical system
must be small enough~\cite{Remark}. In the present case, this
perturbation is the derivative of the potential (see
section~\ref{sect2}) which must thus remain everywhere small enough.
If the derivative were locally diverging for example, KAM theorem
would not apply.  The strategy is therefore to choose a smooth
potential which would develop a singularity under deformation in some
limit. Several choices are possible, depending on which derivative
should be singular~\cite{Remark}, and we will choose the simplest one
with a locally large first derivative. By continuity we expect ( and we
will check) that the threshold of the Aubry transition is strongly
reduced for such modified potentials.  As a matter of fact, there has
been interest over the years in modifying either the interatomic
potentials~\cite{Milchev,Morse, Quapp} or the external periodic
potentials~\cite{PeyrardRemoi1} but the present issue has not been
discussed.

The paper is organized as follows. In section~\ref{sect2}, we
introduce the modified model that is characterized by an amplitude and
a shape parameter that induces locally large derivatives. We study its
Aubry transition (\textit{i.e.} how the pinning is affected) and the
distortions both analytically and numerically in
section~\ref{sect3}. 
In section~\ref{sect4}, we compute
the phason gap that opens up at the transition and compute, more
generally, how the phonon spectrum evolves.

\section{Modification of the Frenkel-Kontorova model} \label{sect2}

The modified classical Frenkel-Kontorova model we consider here reads
\begin{equation}
\label{energy}
W(\{x_n\}) = \frac{1}{2} \sum_n (x_{n+1}-x_n-\mu)^2 + \frac{K}{(2\pi)^2} \sum_n V_{\alpha}(x_n),
\end{equation}
where $x_n$ are the continuous physical variables and $K$, $\alpha$
and $\mu$ some parameters. $x_n$ is typically the position of an atom
$n$ constrained to be along a linear chain.  Here, the first
term of $W$ is a local approximation of an interatomic potential which
has a minimum at $\mu$. $\mu$ can be regarded as tunable, for example
by an external applied force. The second term is a periodic substrate (or interaction)
potential with an amplitude controlled by $K$. Its period is chosen to be 1 (this sets the unit of length with no loss of
generality),
\begin{equation} V_{\alpha}(x+1)=
  V_{\alpha}(x).
\end{equation}
The continuous translation invariance in the absence of the potential,
$x_n \rightarrow x_n+ \phi$ where $\phi$ is any real number, is now
broken by the potential.

We consider the modified potential,
\begin{eqnarray}
  V_{\alpha}(x) &=&  \frac{\cosh \alpha \cos (2\pi x) -1}{\cosh \alpha - \cos (2\pi x)},
\label{pot0} \end{eqnarray}
which depends on a real parameter $\alpha>0$ that controls the shape
(see Fig.~\ref{Potential} and some details in appendix~\ref{appA}). When
$\alpha \rightarrow +\infty$, $V_{\alpha}(x) \rightarrow \cos (2\pi
x)$, giving the standard Frenkel-Kontorova
model~\cite{Aubry}. When $\alpha \rightarrow
0$, the potential is more strongly peaked at integers.  This choice is
interesting because its derivative becomes locally large when $\alpha$
is small. Its maximum is indeed given by (see the appendix)
\begin{equation} V'_{\alpha}(x_0)= \frac{9 \pi}{2 \sqrt{3}}
  \frac{1}{\alpha}, \label{der}
\end{equation}
in the limit of small $\alpha$, whereas the amplitude of the potential
is a constant equal to 2, allowing us to study the effect of a locally
large first derivative. This is the
potential introduced by Peyrard and Remoissenet (up to an irrelevant
constant term)~\cite{PeyrardRemoi1}. It is in contrast with the $\cos (2\pi x)$
choice for which the derivative is bounded by $2\pi$. 

\begin{figure}[h]
  \psfig{file=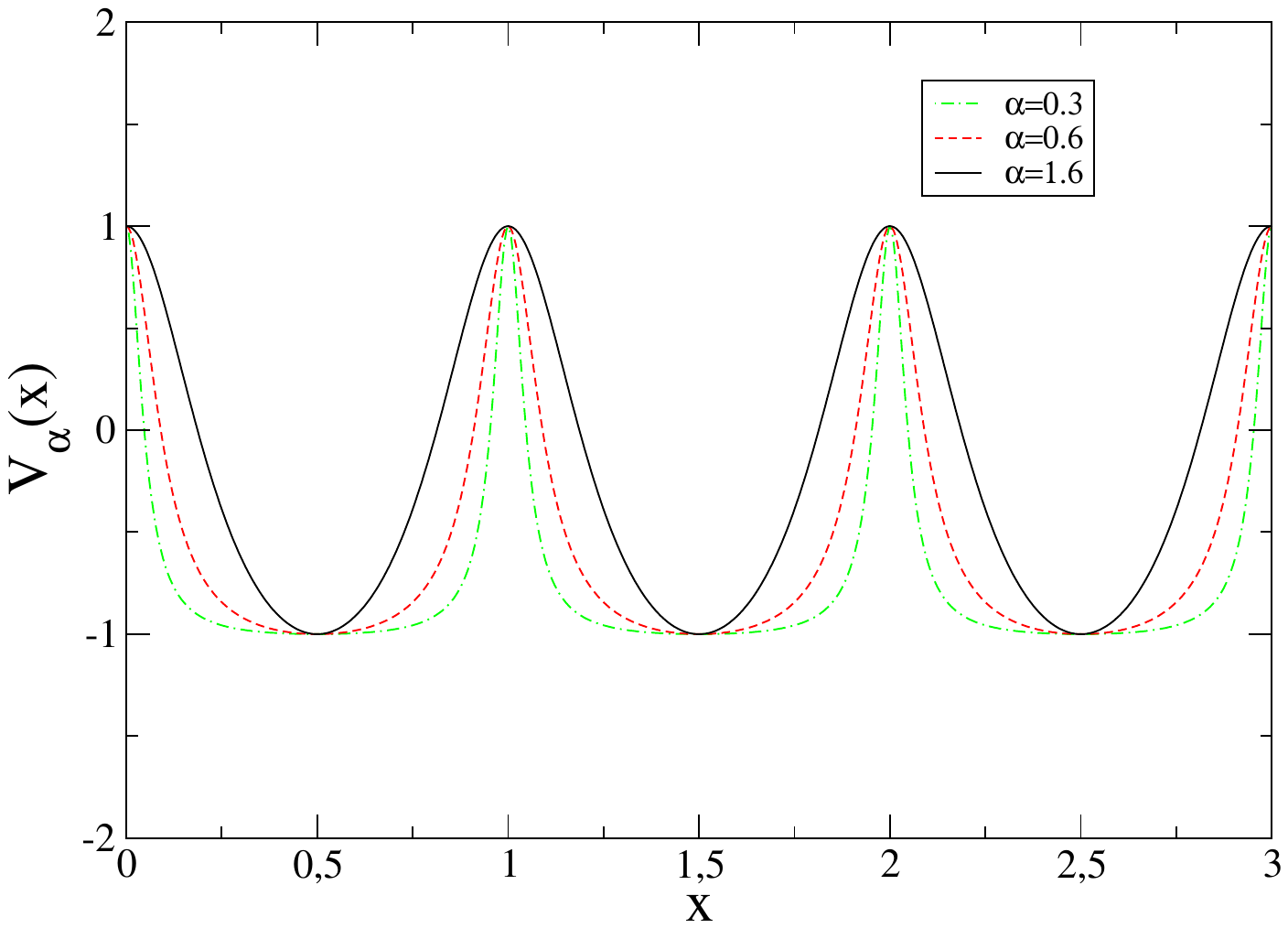,width=10.0cm,angle=-0}
  \caption{Modified periodic potential $V_{\alpha}(x)$ for various $\alpha$. When $\alpha \rightarrow +\infty$, the potential $V_{\alpha}(x) \rightarrow \cos(2\pi x)$, which is the standard Frenkel-Kontorova model. For small $\alpha$, its derivative becomes locally large.}
\label{Potential}
\end{figure}
Its decomposition in Fourier series is given by
\begin{eqnarray}
   V_{\alpha}(x) =  1+ 2 \sinh \alpha \sum_{n=0}^{+\infty} e^{-\alpha n} [\cos (2\pi n x) -1] .
   \label{pot1} \end{eqnarray}
The amplitudes of the successive harmonics decrease exponentially. It
can also be viewed as a train of Lorentzians centered on the integers
(see appendix~\ref{appA}).

We look for the equilibrium configurations $\{x_n\}$ that minimize the
energy $W(\{x_n\})$. They must at least satisfy the equilibrium of forces equation,
\begin{equation}
  \frac{\partial W}{\partial x_n}= 2x_n-x_{n-1}-x_{n+1} + \frac{K}{(2\pi)^2} V_{\alpha}'(x_n)=0.
  \label{firstderivative}
\end{equation}
Note that if $\{x_n \}$ is a solution, $\{x_n+k\}$ where $k$ is any integer is
also a solution and $\{-x_n\}$ too, thanks to the parity of
$V_{\alpha}(x)$, $V_{\alpha}(-x)=V_{\alpha}(x)$. 

Aubry has rewritten the last equation by introducing the bond lengths $\ell_n \equiv x_{n+1}-x_n$, 
\begin{equation}
  \left\{
  \begin{array}{ll}
  x_{n+1} &= x_{n}+\ell_{n},   \\
  \ell_{n+1} &= \ell_{n}+ \frac{K}{(2\pi)^2} V_{\alpha}'(x_{n}+\ell_n). \end{array} \right.
  \label{ds}
\end{equation}
This defines a two-dimensional dynamical system where $n$ is seen as a discrete time. It is known as the standard map when $V_{\alpha}(x)$
reduces to the cosine potential~\cite{Aubry,aubry_ledaeron,mather}. More generally, the nonlinear term of the map 
involves the
derivative of the potential $V_{\alpha}'(x)$, not
the potential itself, and may be large if the derivative is large.
Such maps have chaotic unbounded trajectories when $K$
is large enough but also periodic and quasi-periodic trajectories.

\section{Incommensurate solutions and Aubry transition}
\label{sect3}

\subsection{General background}

In the absence of periodic potential, $K=0$, the solution is simply given by
\begin{equation}
\label{sol00}
x_n=n \ell + \phi,
\end{equation}
where the phase $\phi$ is arbitrary. This is the sliding phase of a trivial integrable model with continuous
translation symmetry.   This state satisfies the balance of forces and becomes the ground state when
$\mu$ equals the lattice constant $\ell$.

When $K \neq 0$, the problem is no longer simple but some exact properties of the ground states are known~\cite{aubry_ledaeron} (see also \cite{bangert} and appendix~\ref{appB}).  
For a general class of models that includes the model we consider here (see appendix~\ref{appB} for the general conditions),
Aubry and Le Daeron have shown that the ground state can be written
\begin{equation}
\label{sol0}
x_n=n \ell + \phi+u_n,
\end{equation}
 with a well-defined lattice constant,
\begin{equation}
  \ell = \lim_{n \rightarrow \infty} \frac{x_n-x_{0}}{n},
\end{equation}
that can take any real value provided that $\mu$ is appropriately tuned. It is thus possible to work at fixed $\ell$.
Importantly, the distortions $\{u_n\}$ are bounded for the ground state and satisfy (see appendix~\ref{appB}):
\begin{equation}
  |x_{n+1}-x_n-\ell| =|u_{n+1}-u_n| \leq 1.
\label{inequality}
\end{equation}
In the ground state, the bond lengths are constrained not to be far from the average $\ell$. 
$\ell$, which is in units of the period of the potential,
 can  be a rational or
irrational number. In the first case,
\begin{equation}
  \ell=\frac{r}{s},
\end{equation}
where $r$ and $s$ are two coprime
integers, one has
\begin{equation}
 x_{n+s}=x_n+r,
\end{equation}
thus $u_n$ is periodic with period $s$, $u_{n+s}=u_n$. In that case,
the ground state is said to be commensurate and has a unit-cell of size $r$ admitting
$s$ atoms at positions $x_n$ ($n=1,\dots, s$).

In the second case, when $\ell$ is an irrational number, the ground state is said to be incommensurate and can be viewed, physically, as a commensurate
solution with a very large period $s$. The distortions can be written
\begin{equation}
  u_n=g(n \ell +\phi),
\end{equation} where $g$ is a periodic function with period 1 which is defined everywhere since  $n \ell+\phi$ takes, modulo 1, all values in $[0,1]$.  The ground state then takes the special form
\begin{equation}
x_n=f(n\ell + \phi), \label{Ans}
\end{equation}
where $f(x)=x+g(x)$ is a strictly increasing function, called the
envelope function. $f$ depends on $\ell$ and on the various model
parameters. The form (\ref{Ans}) is exact for incommensurate
ground states provided that the model fulfills the properties given in
appendix~\ref{appB}.  Importantly, $f$ can be continuous or
discontinuous: the change of regularity with model parameters is 
Aubry's breaking of analyticity (Aubry transition)~\cite{Aubry,aubry_ledaeron}.

For the model we consider, we define $K_c(\alpha)$ as the threshold of the Aubry transition: for $K<K_c(\alpha)$, the ground state is characterized by a \textit{continuous} function $f$.  This is the sliding
phase. For $K>K_c(\alpha)$, $f$ is \textit{discontinuous}: this is the pinned phase. 
The threshold $K_c(\alpha)$ depends on the irrational $\ell$. For the standard Frenkel-Kontorova model ($\alpha \rightarrow \infty$), it is empirically known that the maximal threshold occurs for $\ell=\frac{3-\sqrt{5}}{2}=2-\varphi \approx 0.3819660\dots$
(where $\varphi$ is the golden number) or, equivalently, at $\ell=\varphi-1$, and $K_c(\infty)=0.9716$~\cite{Greene}. In the following we examine the Aubry transition for $\ell=2-\varphi$.

At small $K$, the form of the solution (\ref{Ans}) is coherent with KAM theorem which applies to the dynamical system, Eq.~\ref{ds}. KAM theorem ensures that, if $\ell$ is ``sufficiently" irrational (there is a Diophantian condition~\cite{Remark}), the solution (\ref{sol00})  of the integrable model ($K = 0$) remains a stable trajectory when the nonlinear potential is small enough, up to a change of variable $f$ of the form (\ref{Ans}). In this case, $f$ is a \textit{continuous} bijection.
As a consequence, $x_n~$mod~1 takes all values
in $[0,1]$ just as in the $K=0$ case: the KAM torus (here the circle
$[0,1]$) with that irrational $\ell$ is preserved.  It is known that KAM
theorem does not hold for arbitrary large $K$ although (\ref{Ans}) remains true with a discontinuous $f$.
The threshold at
which all KAM tori cease to exist signals the transition to
``stochasticity''~\cite{Greene} which is also the Aubry
transition for $\ell=2-\varphi$.

\subsection{Small $K$}

 We first mention that the continuous degeneracy of the ground states (\ref{sol00}) at $K=0$ is lifted at first order in perturbation theory for commensurate states but not for incommensurate states.
 We approximate the irrational number $\ell$
by a rational number $\ell=r/s$ (the larger the $s$, the better the
approximation).  At first order, the energy per atom of a unit-cell noted $w^{(1)}$, 
assuming $\mu=\ell$,  is  given by 
\begin{equation}
w^{(1)}=\frac{K}{s (2\pi)^2} \sum_{n=1}^s V_{\alpha}(n \ell+ \phi) = \frac{K}{(2\pi)^2}\frac{\sinh \alpha}{\sinh \alpha s} V_{\alpha s}(s \phi)+C
 \end{equation}
which goes to zero for all $\phi$ when $s \rightarrow +\infty$ ($C$ is a constant independent of $\phi$). It means that the energy
barrier for translating the undistorted state by $\phi$ vanishes for
an ideal incommensurate state at the lowest order (it is remarkable
that it remains true at higher orders as we will see below).  Note
that for a finite $s$, the energy depends on $\phi$ and the
extrema are at $\phi=\frac{m}{2s}$ where $m$ is an integer.

We now solve (\ref{firstderivative}) by a perturbation theory at small $K$.  In order to
determine the distortions $u_n=g(n\ell+\phi)$ one can rewrite the
extremal condition Eq.~(\ref{firstderivative}) as
\begin{equation}
g(x+\ell)+g(x-\ell)-2g(x)=\frac{K}{(2\pi)^2} V_{\alpha}'(x+g(x)),\label{eqA}
\end{equation}
where $x=n\ell+\phi$, and use perturbation theory in $K$ to determine
the periodic function $g$. By using Fourier series, one formally gets,
\begin{equation}
g(x)=  \sum_{p=1}^{+\infty} \frac{A_p}{1-\cos(2\pi p\ell)} \sin 2\pi p x,\label{pert}
\end{equation}
where $A_p$ are some coefficients given, at first and second order in
$K$, by
\begin{eqnarray}
A_p^{(1)} &=& K \frac{\sinh \alpha }{2\pi}  p e^{-\alpha p}, \label{fo} \\
A_p^{(2)} &=& K^2 \frac{\sinh^2 \alpha }{4\pi} \sum_{n \neq p} \frac{n^2(p-n) e^{-\alpha(|n|+|p-n|)}}{1-\cos(2\pi (p-n)\ell)},\label{Ap2}
  \end{eqnarray}
where the last sum excludes all $n$ such that $n-p=ks$ where $k$ is an integer, if $\ell=r/s$.

Does the series Eq.~(\ref{pert}) converge and what is the amplitude of
the distortion?

For rational $\ell=r/s$, $g$ has denominators $1-\cos (2\pi
p\ell)=0$ when $p$ is a multiple of $s$ ($p=ks$), so that (\ref{pert})
is infinite, which means that Eq.~(\ref{eqA}) cannot be satisfied for
all $x$ (or all $\phi$). The only possibility is to choose the special
values $\phi=\frac{m}{2s}$, where $m$ is any integer, which correspond to the extrema of $w^{(1)}$. In this case,
when the denominator vanishes (for $p=ks$), the numerator vanishes
too: $\sin (2 \pi pn \ell+2\pi p \phi)=\sin(2\pi kn + \pi k m
)=0$. Consequently (\ref{eqA}) can be satisfied at these special
points but not for all $x$ or all $\phi$, \textit{i.e.} $g$ is not defined
everywhere, as expected for a commensurate solution.

For irrational $\ell$, however, there are ``small denominators'' but
they never vanish. For $\ell$ ``sufficiently'' irrational (this is
where the Diophantian condition is important), it can be shown that
the first-order series (\ref{pert}) converges, thanks to the
exponential decrease of the harmonics~\cite{Wayne}. The KAM theorem
further ensures that higher-order series in $K$ are convergent as
well, provided that $K$ remains small enough. In this case, $g$ is a
continuous function.
\begin{figure}[t]
      \psfig{file=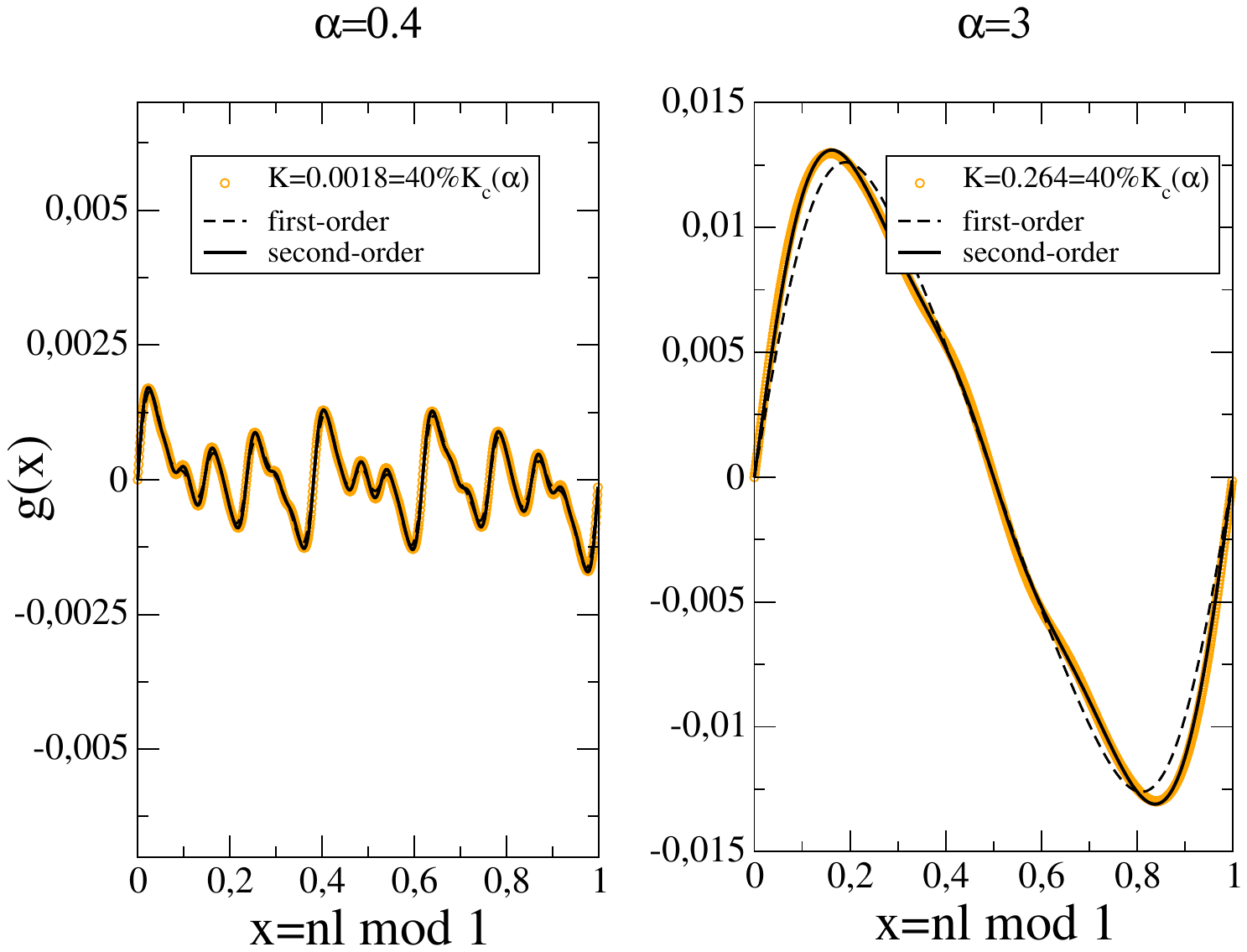,width=10.0cm,angle=-0}
      \caption{
       Continuous envelope function $g$ obtained by numerical  (orange (light gray) points)  and perturbative  (black curves)  [Eq.~\ref{pert}] calculations at small $K=40\%K_c(\alpha)$, for $\alpha=0.4$ (left) and $\alpha=3$ (right). Here $\ell=377/987 \approx 2 -\varphi$.}
\label{env00}
\end{figure}
The amplitude of the distortions $\delta$ defined as \begin{equation} \delta \equiv \mbox{Max}_x~|g(x)| \end{equation} depends on $K$,  $\alpha$ and $\ell$. It is not simply proportional to $K$ but depends on both $\alpha$ and $\ell$ in a complicated way because of the small denominators in (\ref{pert}). 
We now compute the distortions numerically, without relying on perturbation theory.

In Fig.~\ref{env00}, we plot two examples, for two different values of $\alpha$ at
small $K$, of envelope functions $g$ obtained numerically by a gradient descent algorithm which searches a zero of
Eq.~\ref{firstderivative}.  When the algorithm converges (i.e. when the gradient is small), it gives a local mimimum. In the regime of small $K$, there is no metastable states and starting from random configurations always produces the same state characterized by the envelope functions $f$ and $g$ as expected for the ground state. 
We thus obtain in this way, for a rational approximant of $2-\varphi$ (typically $\ell=r/s=377/987$), a periodic configuration $\{x_n\}$ ($n=1,\dots,s$) and distortions $\{u_n\}$ which we plot as a function of $n \ell$~mod 1 to define $g$.
 We then compare them with the perturbative results given by
Eq.~(\ref{pert}) up to second order.
 Perturbation theory is in principle limited to
$K \ll K_c(\alpha)$. By taking $K=40\%K_c(\alpha)$,
we observe in both cases of Fig.~\ref{env00} ($\alpha=0.4$ on
the left and $\alpha=3$ on the right) that perturbation theory is
accurate, especially at second-order. Small deviations are visible and
disappear for smaller values of $K \ll K_c(\alpha)$, or, conversely,
are amplified when $K \rightarrow K_c(\alpha)$. Note that the $y$
scale is not the same, so that the amplitude of the distortions
$\delta$ is much smaller for small $\alpha$ (we have to take smaller
values of $K$ as well as to remain at a fixed distance from the
transition). The shape of the distortions also depends on
$\alpha$. When $\alpha \rightarrow + \infty$, we get the usual Frenkel-Kontorova
model and only the first harmonic $p=1$ is retained in (\ref{pert}). Thus,
$g(x)= \gamma \sin 2\pi x$, $\gamma \equiv \frac{K}{4\pi}
\frac{1}{1-\cos 2\pi \ell}$. This is already close to the result for
$\alpha=3$.  On the other hand, for $\alpha \rightarrow 0$, more and
more harmonics must be included, some of them with a large
denominator but the numerical result remains small (thanks to smaller $K$). This is what is seen in Fig.~\ref{env00} (left).  The
numerical result is thus in agreement with KAM theorem and well
approximated by lowest order perturbation theory.

The distortions are small in this regime and particularly so when
$\alpha$ is small and $K$ is appropriately reduced below the transition.

\subsection{Large $K$} \label{antiinte}

When $K$ increases further, however, the previous perturbation theory
fails. One can consider instead the ``anti-integrable'' limit $K
\rightarrow \infty$~\cite{AA} and do perturbation theory in $1/K$. When $1/K=0$, the solutions of
Eq.~(\ref{firstderivative}) are given by \begin{equation}
  V_{\alpha}'(x_n)=0, \end{equation} so that $x_n$ must be integers or
half-integers. Since there is no determination of $x_{n+1}$ from
$x_n$, any series of integers (or half-integers or a mixing) $\{x_n \}$ is
acceptable, \textit{i.e.} can be random and very ``chaotic''. Among the
solutions, the following one,
\begin{equation}
x_n=[n\ell +\phi] +\frac{1}{2}, \label{sollargeK}
\end{equation}
where $[\dots]$ is the integer part, is special. It is a ground state since all the
atoms are at the bottoms of the potential ($x_n~$mod~1=1/2) 
and its average bond length is $\ell$. It has the expected form given by
Eq.~(\ref{Ans}) with a discontinuous envelope function given by 
$f(x)=[x]+1/2$. Moreover, when $1/K$ is small but nonzero, a perturbative calculation in $1/K$ starting from Eq.~(\ref{sollargeK}) gives the ground state, whereas the other configurations give higher energy metastable states~\cite{AA}. The ground state in perturbation in $1/K$ is thus obtained by 
\begin{equation} f(x)=[x]+\frac{1}{2}+\sum_{n=1}^{\infty} B_n ([x+n\ell ]+[x-n\ell]-2[x]),
\label{anti}
\end{equation} with coefficients $B_n$ vanishing for large $K$. At first and second order in $1/K$, we find
\begin{eqnarray}
  B_n^{(1)} &=& \frac{1}{K} (\frac{1+\cosh \alpha}{\sinh \alpha})^2  \delta_{n,1}, \\
  B_n^{(2)} &=& \frac{1}{K^2} (\frac{1+\cosh \alpha}{\sinh \alpha})^4  \delta_{n,2} -\frac{4}{K} B_n^{(1)}.
\end{eqnarray}
Thus $B_n=B_n^{(1)}+B_n^{(2)}$ (at second order) is a decreasing
function of $n$. The perturbation theory is correct if the prefactor
is small $\frac{1}{K} (\frac{1+\cosh \alpha}{\sinh \alpha})^2 \ll 1$,
which needs, in the limit of small $\alpha$, that $K \alpha^2/4 \gg
1$.

In Eq.~(\ref{anti}), the periodic function
$[x+a]+[x-a]-2[x]$ is discontinuous at points $na$, for all $n$.
Therefore $f$ is discontinuous at each point $x=
\pm n\ell~$mod~1, \textit{i.e.} everywhere (since  $n\ell~$mod~1 is dense in $[0,1]$ for $\ell$ irrational), with discontinuities
that are functions of $B_n$. 
\begin{figure}[t]
      \psfig{file=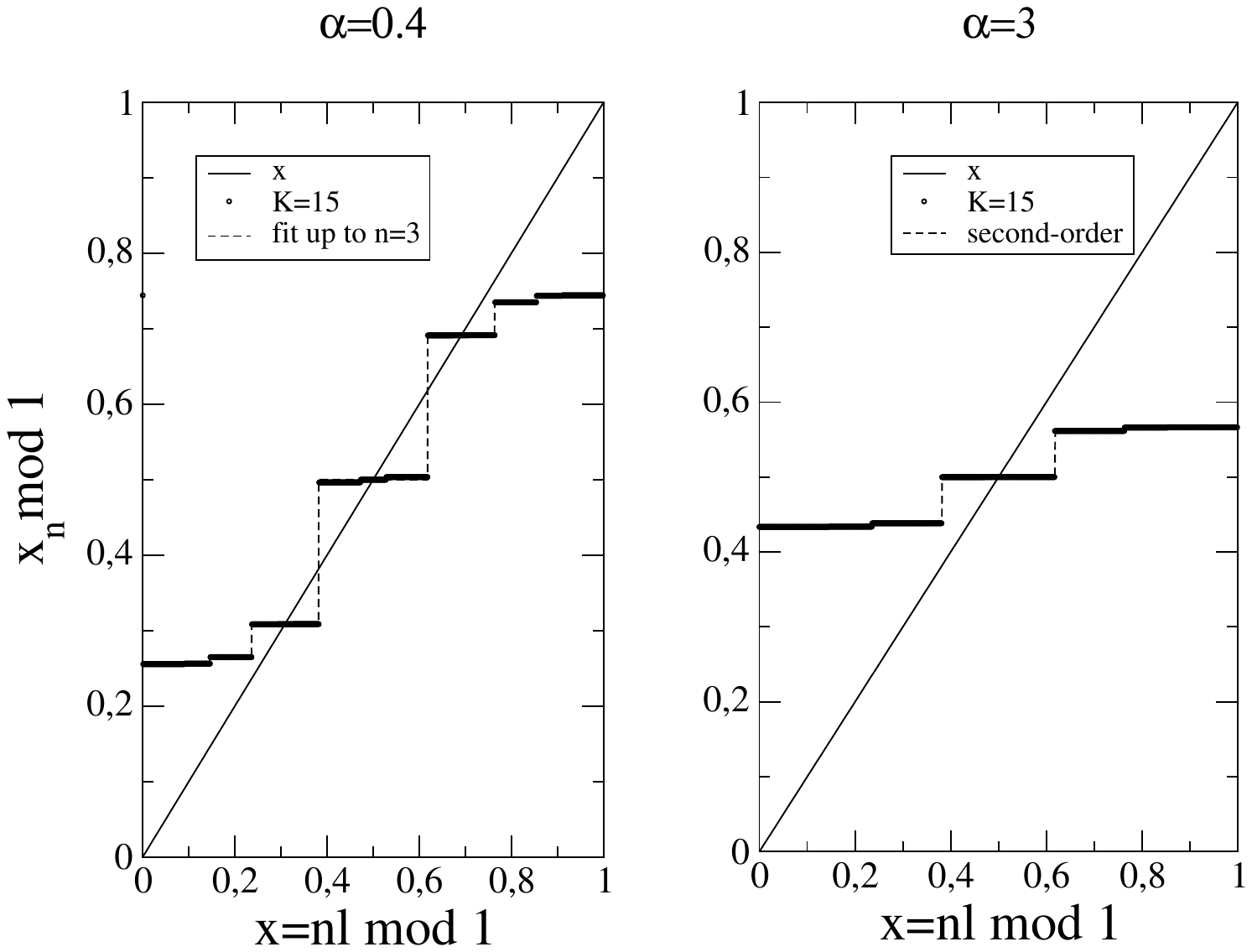,width=10.0cm,angle=-0}
 \caption{Discontinuous envelope function $f$ obtained by numerical and perturbative [Eq.~(\ref{anti})] calculations for large $K=15$ and for  $\alpha=0.4$ (left) and $\alpha=3$ (right). Here $\ell=377/987 \approx 2 -\varphi$. Note that it is impossible for these parameter values to distinguish the numerical result from the perturbative calculation. }
   \label{envanti}
\end{figure}

In Fig.~\ref{envanti}, the points are the atomic positions $x_n$~mod~1
computed numerically for large $K$ and plotted as a function of
$n\ell$~mod~1. The gradient descent, started with a configuration sufficiently close to (\ref{anti}), converges to the ground state.
 We thus obtain the function $f$ which is increasing and discontinuous, as expected for a ground state in this regime. 
For $\alpha=3$ (Fig.~\ref{envanti}, right), the second order result  from Eq.~\ref{anti}
is shown as well (dashed line) and is a good approximation of the
numerical result with main discontinuities at zero (obtained at zeroth
order), $\pm \ell$ (first-order) and $\pm 2 \ell~$mod~$1$ (second
order). The other discontinuities obtained numerically are not reproduced at this
order. The largest discontinuity at zero means that atoms avoid the
maxima of the potential. For $\alpha=0.4$ (Fig.~\ref{envanti}, left), the lowest
order perturbation theory already fails even for $K=15$ because the
prefactor is of order 1. The result shown by a dashed line is a fit
that uses (\ref{anti}) and fitting parameters $B_n$ up to $n=3$,
reproducing the main three discontinuities. The form of the solution
(\ref{anti}) seems to remain accurate even though the parameters $B_n$
are no longer given by perturbation theory. In both cases, we see that
the distortions (departure from the $y=x$ line) are
strong, \begin{equation} \delta \sim O(1).
  \end{equation}
The interpolation between the two previous regimes $K \rightarrow 0$ and $K \rightarrow \infty$ is through the Aubry transition.

\subsection{Aubry transition} \label{pinning}

\begin{figure}[h]
  \psfig{file=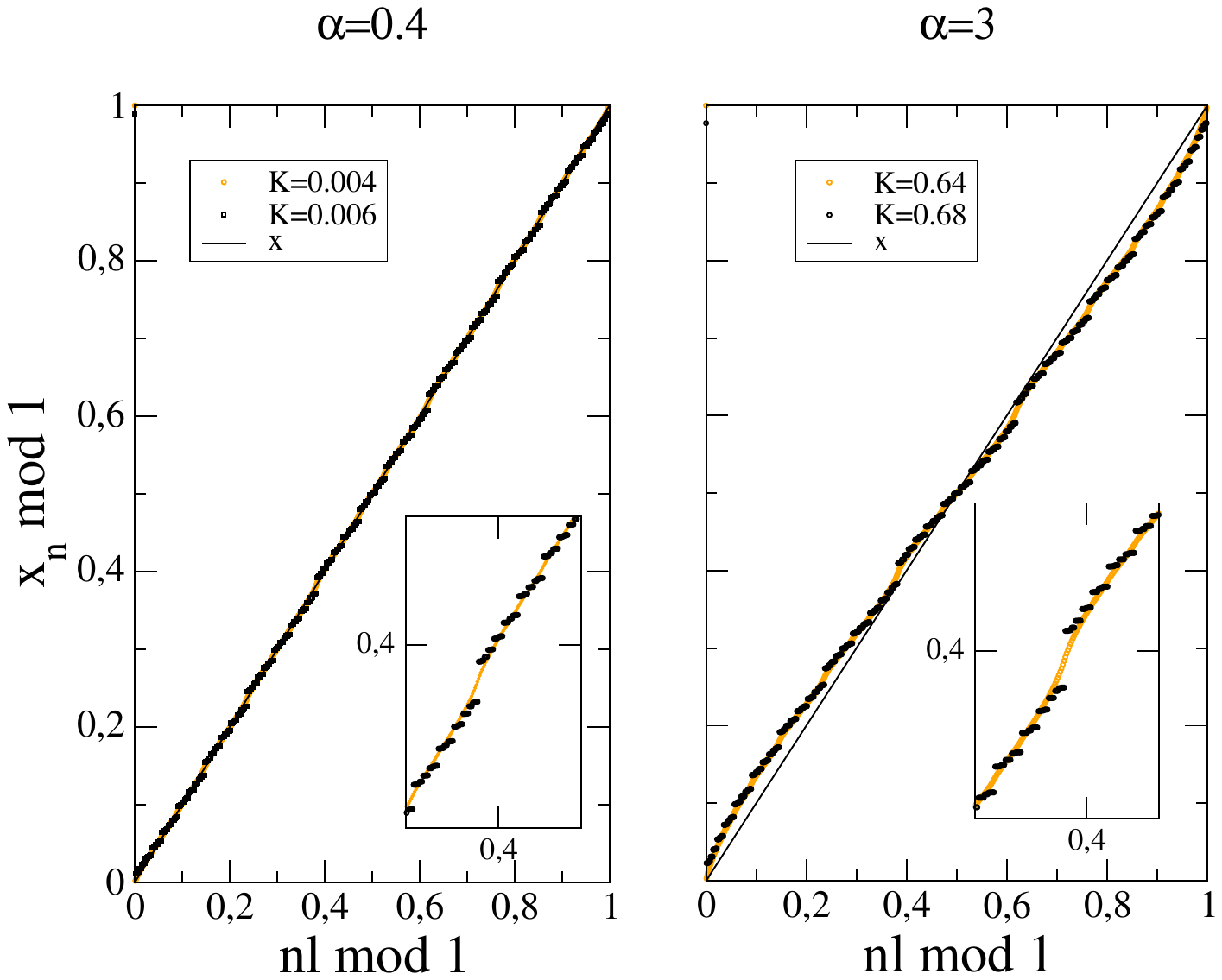,width=10.0cm,angle=-0}
        \psfig{file=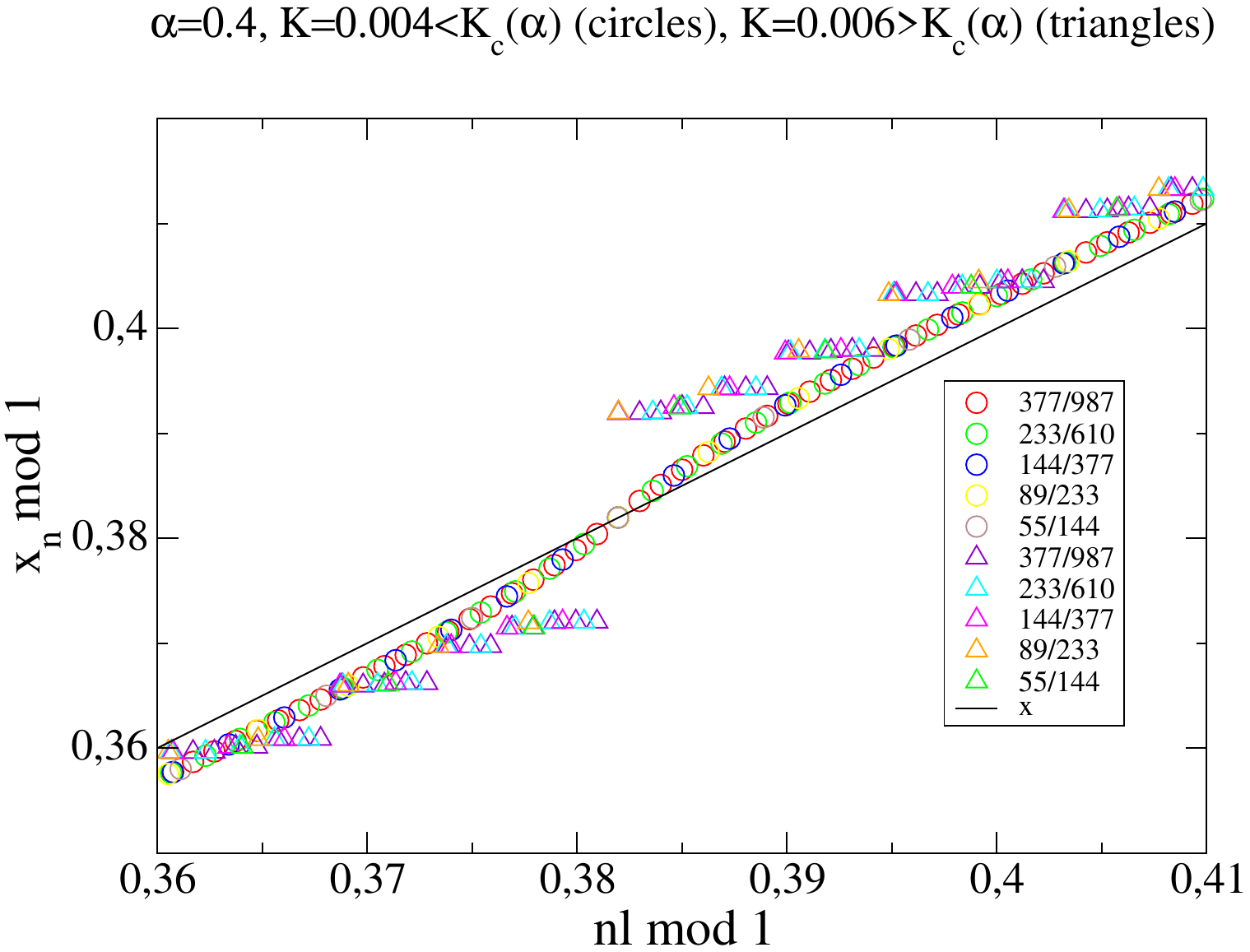,width=9.5cm,angle=-0}
 \caption{Envelope functions $f$ for $\alpha=0.4$ (top left) and $\alpha=3$ (top right) (zoom in the insets): the  orange  (light gray) points form a continuous curve for $K<K_c(\alpha)$, whereas the black points form a discontinuous one for $K>K_c(\alpha)$ . Here $\ell=r/s=377/987 \approx 2-\varphi$. Bottom: two examples of converged $f$ just below (circles) and above (triangles) the Aubry transition for a series of fractions converging to $2-\varphi$. The undistorted result $K=0$ is given by $f(x)=x$.}
\label{env0}
\end{figure}

\begin{figure}[h]
  \psfig{file=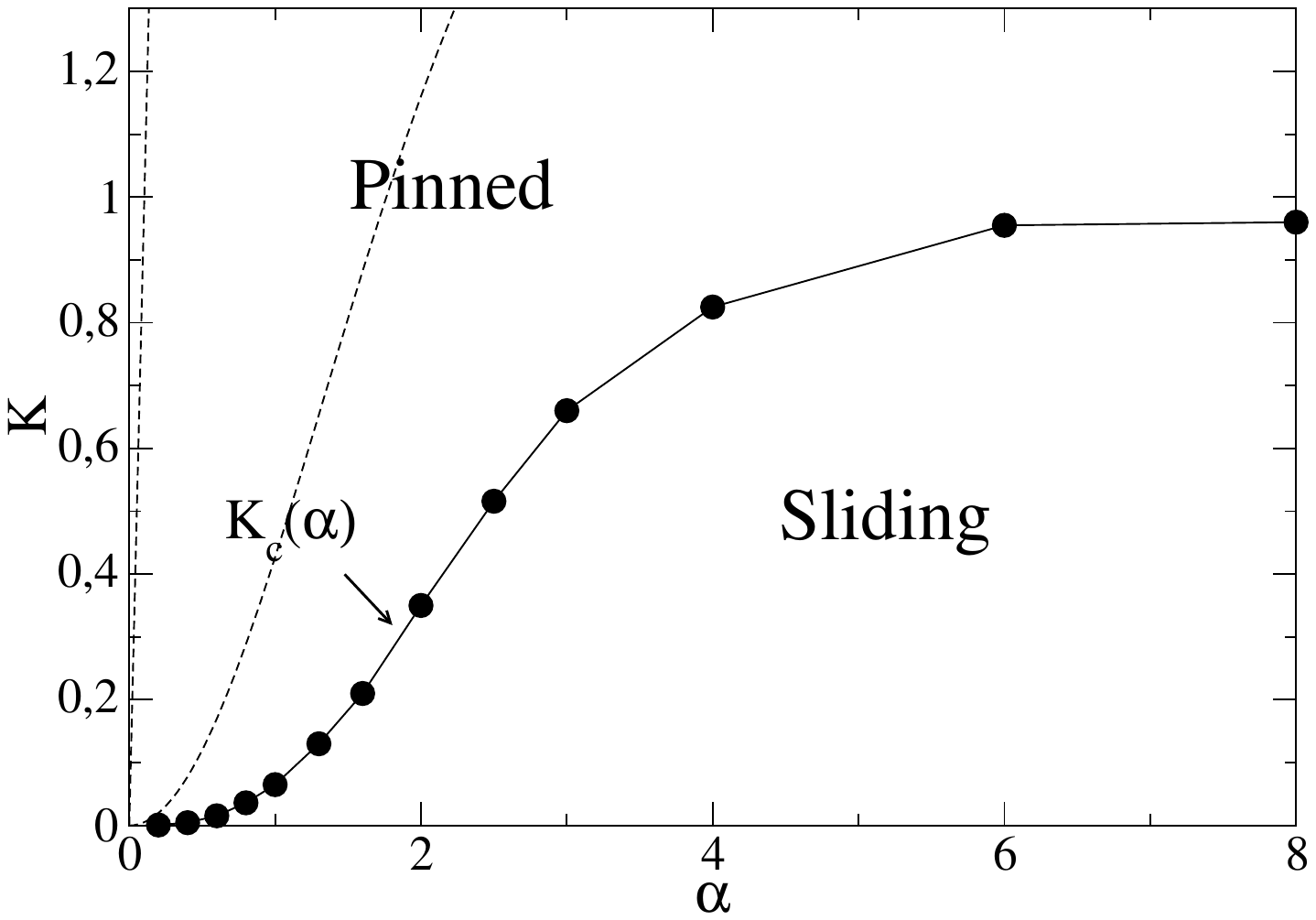,width=10.0cm,angle=-0}
 \caption{Phase diagram with the sliding and pinned phase separated by the Aubry transition for incommensurate $\ell=2-\varphi$. The two dashed lines are crude analytical upper bounds of $K_c(\alpha)$ given by Eqs.~(\ref{bf}) and (\ref{approx2}). } 
\label{pd}
\end{figure}

The simplest way to show numerically the existence of an Aubry
transition is to follow the discontinuities of the envelope
functions~\cite{AubryQuemerais}.  In Fig.~\ref{env0}, we
show, as above, the numerically computed envelope function $f$. Since
$K$ is reduced compared with the results of Fig.~\ref{envanti}, the
distortions are reduced and $f(x)$ gets closer to $f(x)=x$.  In each
figure (top), two examples of values of $K$ close to the threshold of
the Aubry transition, $K_c(\alpha)$, are given.  The  orange (light gray)  points
($K<K_c(\alpha)$) are the points of a continuous function (see insets
for more clarity), just as in Fig.~\ref{env00}, and the black points
($K>K_c(\alpha)$) are that of a discontinuous function as in
Fig.~\ref{envanti}.  We observe that the discontinuities close
continuously so that the transition is a second-order
transition. Furthermore, we have checked the convergence of the
envelope function for successive rational approximants of
$2-\varphi$. Two examples, one for $K<K_c(\alpha)$ and one for
$K>K_c(\alpha)$, are given in Fig.~\ref{env0} (bottom). Although,
strictly speaking, there is no Aubry transition in commensurate
systems, using such large values of $s$ ensures a sharp change of
continuity as a function of $K$.  By locating the value of $K$ for
which the discontinuities close, we extract $K_c(\alpha)$ and the
phase diagram  showing  the sliding phase $K<K_c(\alpha)$ and the pinned
phase $K>K_c(\alpha)$ (Fig.~\ref{pd}). Note that $K_c(\alpha)$
vanishes when $\alpha \rightarrow 0$.

This can be simply understood by adapting an earlier argument~\cite{aubry_minor} on the equilibrium of
forces, Eq.~(\ref{firstderivative}).  From the existence of a bound on the distortions expressed by Eq.~(\ref{inequality}), one obtains 
 for the first part of Eq.~(\ref{firstderivative}) that
$ \vert 2x_n - x_{n+1}-x_{n-1} \vert \leq 2$.  In order to satisfy Eq.~(\ref{firstderivative}), its second part given by $ \frac{K}{(2\pi)^2} V_{\alpha}'(x_n)$ cannot be arbitrary large. For irrational $\ell$ and
sliding solution, $x_n$~mod~1 takes all values in $[0,1]$ so that the maximum of $V'_{\alpha}(x)$, noted $V'_m$, is necessarily reached. Therefore, if
$
\frac{K}{(2\pi)^2} V'_m > 2$,
some atoms in the sliding phase cannot be at equilibrium.
In
the limit of small $\alpha$, given the expression of the maximum of
the derivative (\ref{der}), we get that,  for
\begin{equation}
  K > \frac{8\pi^2}{V'_m}= \frac{16 \pi \sqrt{3}}{9} \alpha \hspace{.5cm} (\alpha \rightarrow 0), \label{bf}
\end{equation}
it is impossible to maintain the balance of forces in the sliding phase.
In particular, when $\alpha=0$ 
the right-hand side vanishes so that \begin{equation}
  K_c(0)=0. \end{equation} At the limit $\alpha=0$ the potential is
discontinuous at $x=0$ and the KAM torus is destroyed at that
point.
This qualitative argument confirms that in the limit
of small $\alpha$, the pinned phase should be favored at small $K$.
Importantly, one sees that if the derivative of the potential $V'_m$
is \textit{somewhere} strong enough, the sliding phase no longer exists
because the atoms that experience that strong force (some necessarily do in
the sliding phase) cannot be at equilibrium: the Aubry transition
threshold of the pinned state is reduced.
The bound in (\ref{bf}) is in fact a very crude estimate of
$K_c(\alpha)$ (see the steep dashed line in Fig.~\ref{pd}). A
better bound represented by the dashed curve in Fig.~\ref{pd} will be given in
section~\ref{sect4}. 


\begin{figure}[h]
   \psfig{file=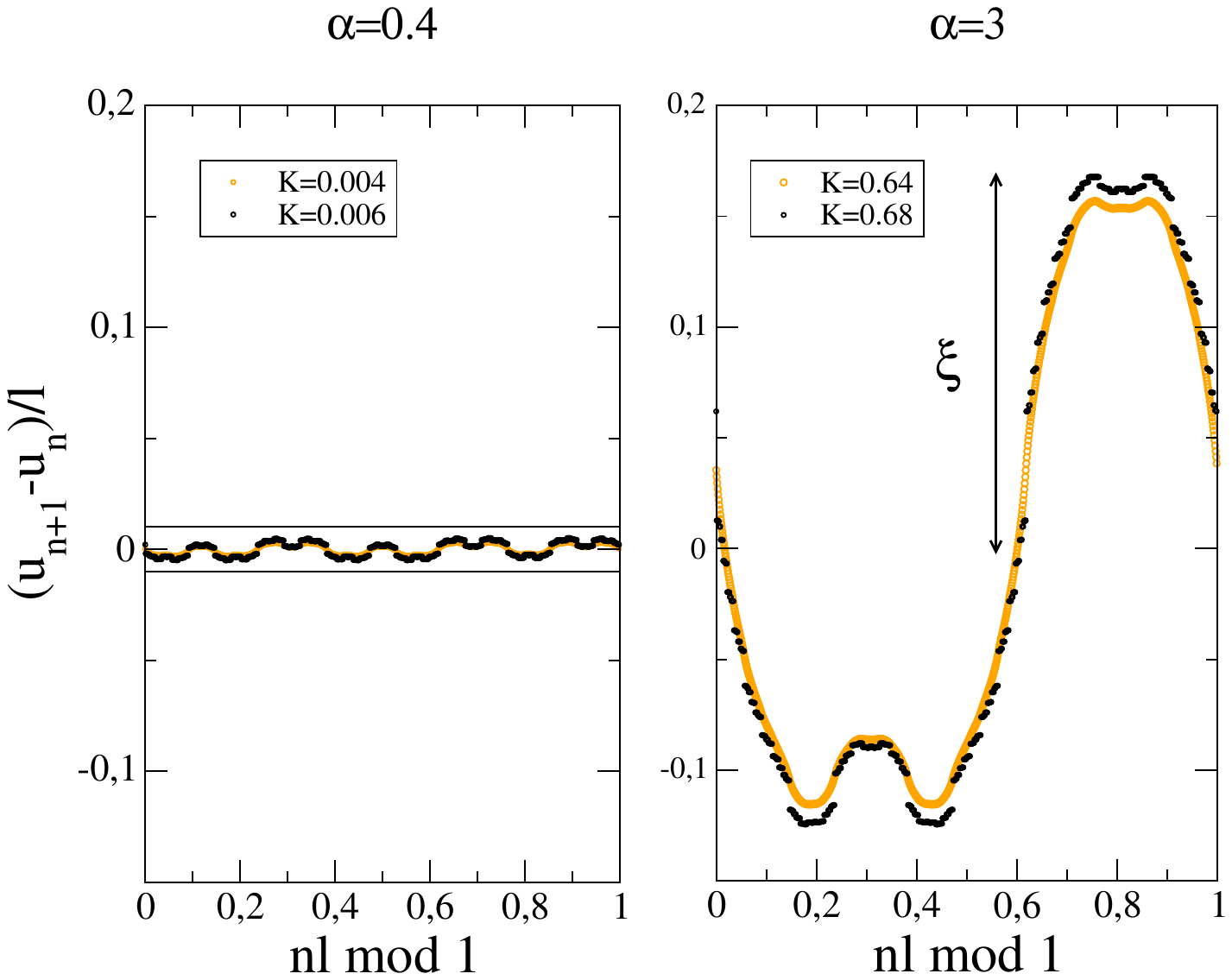,width=10.0cm,angle=-0}
 \caption{Envelope functions $h$ for the normalized bond lengths for $\alpha=0.4$ (left) and $\alpha=3$ (right): the  orange (light gray) points form a continuous curve for $K<K_c(\alpha)$, whereas the black points form a discontinuous one for $K>K_c(\alpha)$ . The amplitude is noted $\xi$. The horizontal bars on the left show $\pm$ 1\%.}
\label{env1}
\end{figure}

\begin{figure}[h]
  \psfig{file=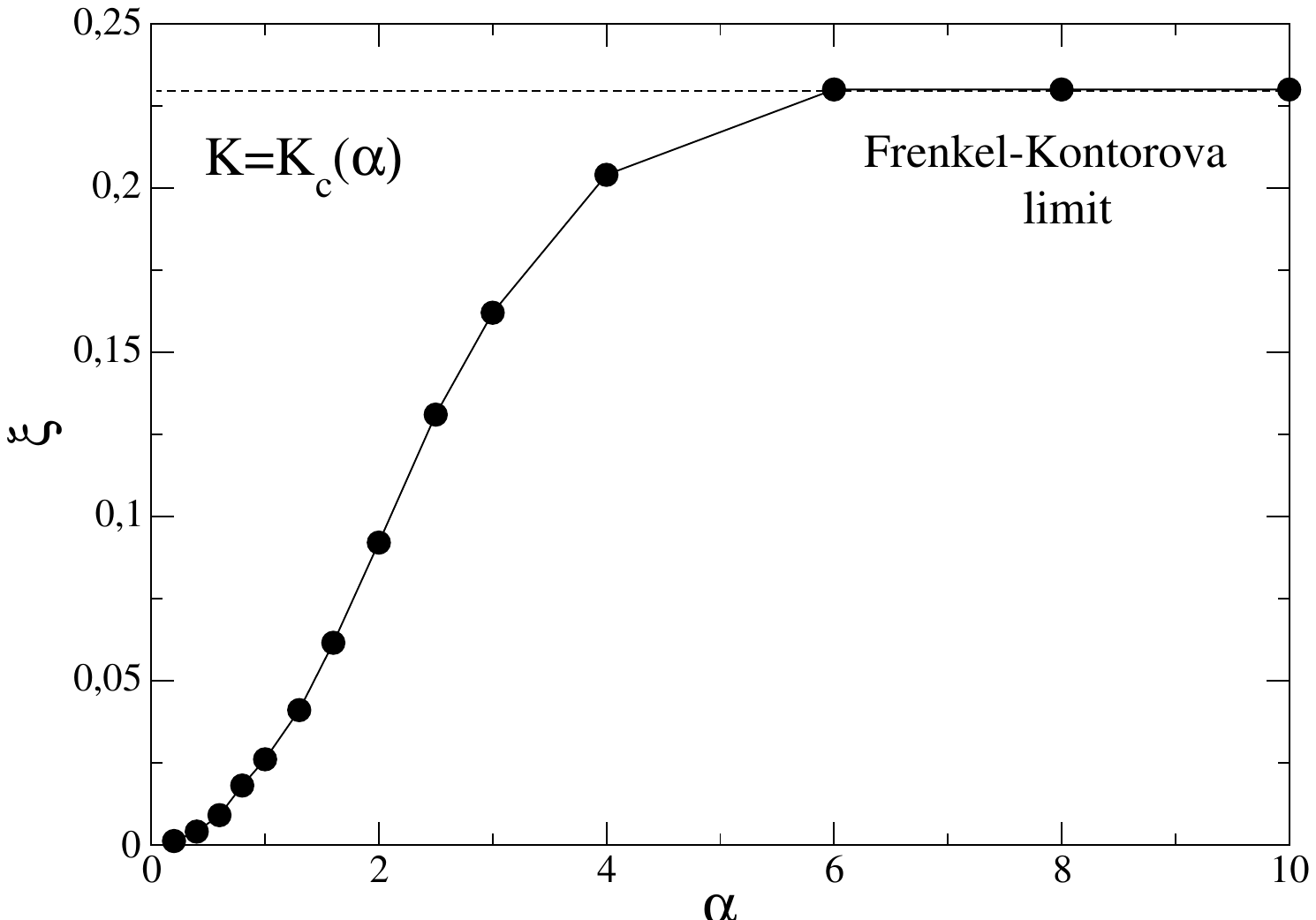,width=10cm,angle=-0}
 \caption{Maximal distortion $\xi$ (see definition in Fig.~\ref{env1}) as a function of the potential shape parameter $\alpha$, at the Aubry transition. At small $\alpha$, $\xi$ can be arbitrarily small and reaches 23\% (dashed line) at large $\alpha$   (Frenkel-Kontorova limit).}
\label{env3}
\end{figure}

In order to get some quantitative insights into how large the
distortions can be near and at the Aubry transition when the phase is pinned, we also compute numerically
the bond lengths,
\begin{eqnarray}
  \ell_n &=& \ell + (u_{n+1}-u_n).
\end{eqnarray}
Recall that the average bond length is $\frac{1}{s} \sum_{n=1}^s
\ell_n=\ell$ so that $(u_{n+1}-u_n)/\ell$ measures the amplitude of
the distortion with respect to the average. We define another envelope function $h$ by $(u_{n+1}-u_n)/\ell=h(n\ell+\phi)$. Its amplitude
\begin{equation} \xi \equiv
  \mbox{Max}_x~|h(x)|,
\end{equation}
gives an idea of how much distorted the structure is.
In Fig.~\ref{env1}, we
show $(u_{n+1}-u_n)/\ell$  as a function of $n \ell~$mod~1 (\textit{i.e.} $h$) for two values of $K$, one for  $K <
K_c(\alpha)$ (continuous curves), the other for $K>K_c(\alpha)$ (discontinuous curve).
  For large
values of $\alpha$ ($\alpha=3$ in Fig.~\ref{env1}, right), the amplitude is
large. On the contrary, for small $\alpha$ ($\alpha=0.4$ in Fig.~\ref{env1}, left), we observe that the
distortions are well within $1~\%$ of the bond length (the two
horizontal lines correspond to $\pm 1~\%$).  The amplitude $\xi$ is reported
in Fig.~\ref{env3} as a function of $\alpha$ at $K=K_c(\alpha)$.  For
large $\alpha$, the maximal distortion at the transition is 23\%
(Frenkel-Kontorova limit). For smaller values of $\alpha$, $\xi$ can
be arbitrary small.  It can also be seen qualitatively in
Fig.~\ref{env0} that for small $\alpha$ (left), the distortions are
very small and $f$ is very close to the undistorted result $f(x)=x$,
whatever the values of $K$ across and near the Aubry transition.

The important conclusion is that it is not necessary to have large
distortions to have an Aubry transition or be in the pinned phase.  In
particular, if $\alpha=0$, so that the potential is discontinuous,
$K_c(0)=0$ as we have shown above: the system is immediately in the pinned phase. Yet the
potential is flat almost everywhere so that there is no distortion. This
extreme situation remains somehow valid at small $\alpha$, as shown here:  it is
replaced by a pinned phase with small distortions.
 

\section{Stability and phonon spectrum}
\label{sect4}

We now examine the stability of the solution and the phonon spectrum,
in particular its gap which vanishes at the Aubry transition, defining
the zero-energy phason mode of the continuous ground state manifold
for $K<K_c(\alpha)$.  For this, we add the kinetic energy of the
atoms
\begin{equation}
  H = \frac{1}{2} \sum_n \dot{x}_n^2 +  W(\{x_n\}),
\end{equation}
and write
\begin{equation}
  x_n=x_n^{eq}+\epsilon_n,
\end{equation}
where $x_n^{eq}=n\ell +u_n$ is the equilibrium position in the ground
state previously obtained and $\epsilon_n$ a sufficiently small
deviation to expand the energy:
\begin{eqnarray}
W(\{x_i\})= W(\{x_i^{eq}\})+\frac{1}{2} \sum_{n,m} \frac{\partial^2 W}{\partial x_n \partial x_m} \epsilon_n \epsilon_m. \label{mat}
\end{eqnarray}
The nonzero partial derivatives are given by
 \begin{eqnarray}
 \frac{\partial^2 W}{\partial x_n^2}  &=& 2 + \frac{K}{(2\pi)^2} V_{\alpha}''(x_n^{eq}), \\
 \frac{\partial^2 W}{\partial x_n \partial x_{n \pm 1}}  &=& -1,
\end{eqnarray}
where $V''_{\alpha}(x)$ is given in appendix~\ref{appA}.  Note that for the equilibrium phase to be a minimum of the
energy, the matrix on the right-hand-side of (\ref{mat}) must be
definite positive. The (Sylvester) criterion implies in particular
that all diagonal elements must be positive~\cite{aubry_minor}. In the
sliding phase, all values of $V''_{\alpha}(x_n)$ are attained, in
particular its minimum $V''_{\alpha}(0)<0$. When $K$ increases, $ 2 +
\frac{K}{(2\pi)^2} V_{\alpha}''(0)$ becomes negative and the matrix is
no longer definite positive (the sliding phase is unstable), \textit{i.e.} when
\begin{eqnarray}
K> \frac{2(2\pi)^2}{|V''_{\alpha}(0)|}=2 \frac{\cosh \alpha -1}{\cosh \alpha +1} .  \label{approx2}
\end{eqnarray}
This analytical bound is a
crude approximation but is in agreement with the numerical result giving $K_c(\alpha)$ (see
the dashed curve in Fig.~\ref{pd} for a comparison). It could be
refined by using higher-order minors~\cite{aubry_minor} but it is not
necessary here.  We find again that when $\alpha \ll 1$, the sliding phase must be unstable above $K \sim
\alpha^2/2$ which is small. 

To compute the phonons we now assume a commensurate state with
$\ell=r/s$ and that $\epsilon_n=\bar{\epsilon}_n e^{i (kn- \omega_k
  t)}$ (the amplitude $\bar{\epsilon}_n$ is periodic with period $s$)
and obtain an $s\times s$ matrix:
\begin{equation}
  M_{n,m}=-\omega_k^2 \delta_{n,m}+ \frac{\partial^2 W}{\partial x_n \partial x_m}.
\end{equation}
The matrix $M$ has also the nonzero end points $ \frac{\partial^2
  W}{\partial x_s \partial x_{1}} = -e^{\pm iks} $ for $\ell \neq
0,1/2$.  The diagonalization of $M$ gives the phonon energies
$\omega(k)$.  For the integrable point $K=0$, the spectrum is simply
given by the standard expression:
\begin{equation}
\omega(k)=2|\sin \frac{k}{2}|.\label{s}
\end{equation}
In the opposite anti-integrable limit $K \rightarrow + \infty$, 
the atoms are all at the bottoms of the potential,
$V''_{\alpha}(x_n)=V''_{\alpha}(1/2)$ and the dispersion relation is
\begin{equation}
\omega(k)=\sqrt{\Delta^2+4\sin^2 \frac{k}{2}},\label{sKlarge}
\end{equation}
with gap $\Delta$ given by
$\Delta^2=\frac{K}{(2\pi)^2}V''_{\alpha}(1/2)=K \frac{\sinh^2
  \alpha}{(1+\cosh \alpha)^2}$ (see (\ref{sd2})). Note that
the anti-integrable limit applies precisely when $\Delta \gg 1$ (see
section~\ref{antiinte}).  In this case, the spectrum consists  of a
single (gapped) band.

\begin{figure}[h]
  \psfig{file=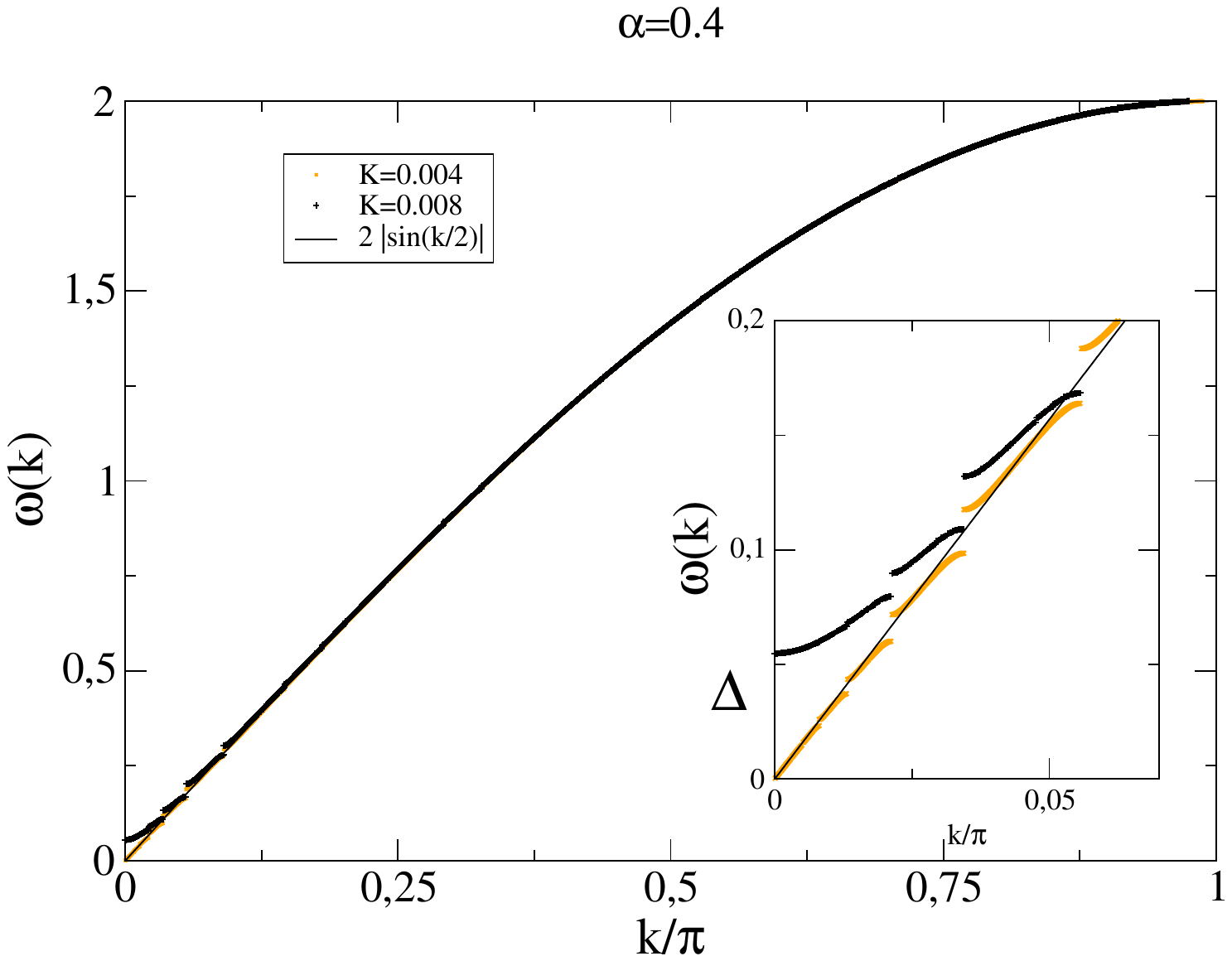,width=10.0cm,angle=-0}
    \psfig{file=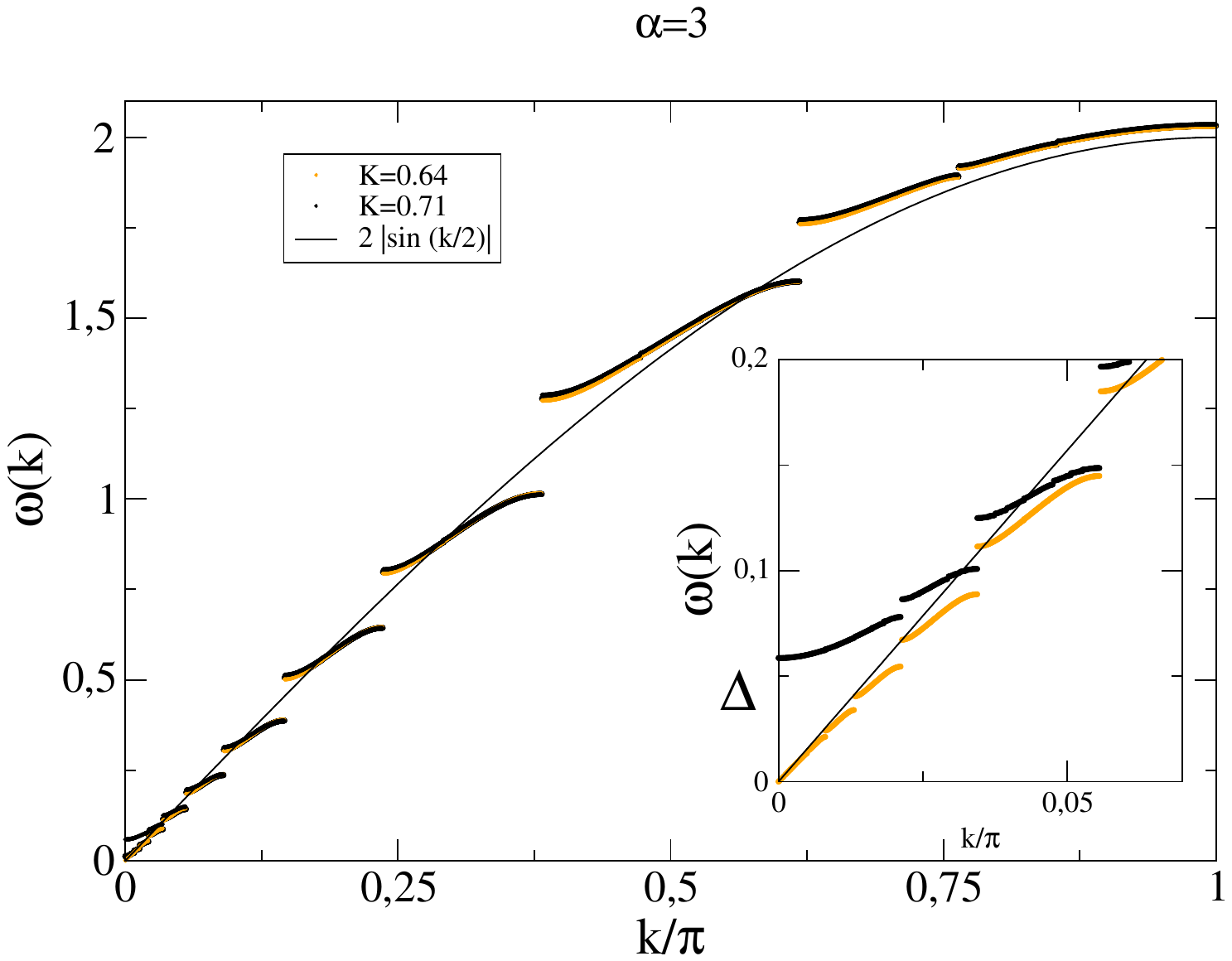,width=10.0cm,angle=-0}
 \caption{Phonon spectrum for $\alpha=0.4$ and $\alpha=3$ (zoom in the insets). For $K<K_c(\alpha)$ (orange (light gray) curves)  the spectrum has no gap at $k=0$ (phason mode). For $K>K_c(\alpha)$ the spectrum has a gap at $k=0$ (black curves). Here $\ell=377/987$.}
\label{phs}
\end{figure}

In between, for $K\neq 0$, the spectrum $\omega(k)$ is computed
numerically. Two examples are given in Fig.~\ref{phs} for
$\ell=377/987 \approx 2-\varphi$ and two different values of $\alpha$. $\omega(k)$ is represented in the half extended Brillouin zone,
\textit{i.e.} for $k$ in $[0,\pi]$.

In both cases the spectra of the sliding phases ($K<K_c(\alpha)$,  orange (light gray)  curves) have
no zero energy gap. This is the consequence of the existence of a
continuous manifold of ground states. The associated zero-energy mode
is the 'phason', which becomes gapped in the pinned phase above the
Aubry transition $K>K_c(\alpha)$ (black curves).  Note that the parameters $K$ in gapped cases
are chosen so that the gap is the same in both figures (see
the black curves in the insets). The values of $K$ (0.008 and 0.71) differ by almost two orders of magnitude.
Measuring experimentally a certain gap is therefore not sufficient
to tell what the amplitude $K$ of the potential is. 

For small $\alpha$ (top), the distortions and the values of $K$ at the
Aubry transition are small, so that the spectrum is closer to the standard phonon spectrum
(\ref{s}).  For large $\alpha$ (bottom), there are much larger gaps at
higher energies.  Since the periodic potential mixes modes with $k$
and $k \pm 2\pi p\ell$ (where $p$ is an integer),  one expects gaps at
every level crossing. One sees that for small $\alpha$ the high-energy
gaps are all very small while they are large when $\alpha$ is
large. This simply reflects the strength of the distortions. It makes
a qualitative difference which may help to distinguish experimentally
between strongly nonlinear potentials (small $\alpha$) and harmonic
potentials (large $\alpha$).

\section{Conclusion}

The distortions of incommensurately modulated phases are not
necessarily as strong as what the Frenkel-Kontorova model or
charge-density wave models suggest when pinning occurs. When the
smoothness conditions of the nonlinear perturbation are progressively
suppressed, \textit{i.e.} here when the derivative of the potential (the
mechanical force) becomes locally strong enough, the pinning threshold $K_c(\alpha)$ is reduced.
 This is  coherent with the fact that KAM theorem  no longer
applies for potentials with some singularities.  At the same time, the
distortions can be weak if the potential is flatter in large portions of
space. This is what we have provided evidence for by considering
a simple modified potential in which these two regions coexist, a situation
that does not occur in the standard Frenkel-Kontorova model where the
derivative of the potential is bounded.  The two effects
-pinning and distortions- are therefore not necessarily related.  This
opens a wide range of applicability of the Aubry transition since distortions need not to be
large in incommensurate pinned phases.

We therefore emphasize that  observing experimentally small incommensurate distortions $\xi$ (as in charge-density waves) does
not generally imply that the phase should be sliding. It is only true for
the standard Frenkel-Kontorova model. It does not imply either that
perturbation theory, which leads to the sliding phase, is applicable
because it is highly ``resonant'' due to the small denominators. The
small parameter of the perturbation theory is $K/K_c(\alpha)$, not
$\xi$.   For the standard Frenkel-Kontorova model, we have simultaneously
$\xi \ll 1$ and $K/K_c \ll~1$. However, in general cases, $\xi$ may be small while $K/K_c(\alpha)$ remains large, so that perturbation theory does not apply. 

\appendix

\section{Modified potential}
\label{appA}

The modified potential considered here depends on a real parameter $\alpha$ and is written in three different forms:
\begin{eqnarray}
  V_{\alpha}(x) &=&  \frac{\cosh \alpha \cos (2\pi x) -1}{\cosh \alpha - \cos (2\pi x)} \label{pot00} \\
  &=&  1+ 2 \sinh \alpha \sum_{n=0}^{+\infty} e^{-\alpha n} [\cos (2\pi n x) -1]    \label{pot1} \\
  &=&  \sum_{n=-\infty}^{+\infty} \frac{ 2 \alpha \sinh \alpha }{[2 \pi (x-n)]^2+ \alpha^2} -\cosh \alpha. \label{pot2}
 \end{eqnarray}
It is easy to prove these equalities. For example, starting from (\ref{pot2}) and using Poisson's formula for $h(x) \equiv \frac{\alpha}{4 \pi^2 x^2+ \alpha^2}$,  
\begin{eqnarray}
  \sum_{n=-\infty}^{+\infty} h(x-n) = \sum_{p=-\infty}^{\infty} \hat{c}_p e^{2ip\pi x} \end{eqnarray}
with
\begin{eqnarray}
   \hat{c}_p=\int_{-\infty}^{+\infty} h(x) e^{-2i\pi p x}dx=\frac{1}{2} e^{-|p|\alpha},
  \end{eqnarray}
we find (\ref{pot1}). Then, by resummation of the Fourier series (\ref{pot1}), we get (\ref{pot00}).

The first and second derivatives of the potential are given by
\begin{eqnarray}
  V_{\alpha}'(x) &=& - 2\pi \frac{\sinh^2 \alpha \sin (2\pi x) }{[\cosh \alpha - \cos (2\pi x)]^2}.
  \label{deriv} \end{eqnarray}
\begin{equation}
  V_{\alpha}''(x)/(2\pi)^2=\sinh^2 \alpha \frac{1+\sin^2 (2\pi x)-\cosh \alpha \cos (2\pi x)}{[\cosh \alpha-\cos (2\pi x)]^3}.
\end{equation}
The first derivative has a maximum which, for small $\alpha$, lies at $x_0=\alpha/(2\pi \sqrt{3})$, so that  $V'_{\alpha}(x_0)=\frac{9 \pi}{2 \sqrt{3}} \frac{1}{\alpha}$ which is large when $\alpha $ is small.
We also have
\begin{eqnarray}
  V_{\alpha}''(0)/(2\pi)^2 &=& -\frac{\cosh \alpha +1}{\cosh \alpha -1} \label{sd1} \\
  V_{\alpha}''(1/2)/(2\pi)^2 &=& \frac{\sinh^2 \alpha }{(1+\cosh \alpha)^2} . \label{sd2}
\end{eqnarray}

\section{Some exact results}
\label{appB}

The energy given by Eq.~(\ref{energy}) can be written as:
\begin{equation}
\label{newenergy}
W(\{x_n\}) = \sum_n H(x_{n+1},x_n)-\mu \sum_n \left(x_{n+1}-x_n \right),
\end{equation}
with the  twice  differentiable function $H$:
\begin{equation}
\label{H}
H(x,y)=\frac{1}{2} \left[(x-y)^2 + V_{\alpha}(x)+V_{\alpha}(y) + \mu^2 \right].
\end{equation}
Because the function $H$ satisfies a convexity condition,
\begin{equation}
\label{convexity}
\frac{\partial H}{\partial x \partial y} =-1<0,
\end{equation}
together with a condition of periodicity,
\begin{equation}
\label{biperiodicity}
H(x+1,y+1)=H(x,y),
\end{equation}
the model (\ref{energy}) belongs to the class of Frenkel-Kontorova models studied in \cite{aubry_ledaeron}. Slightly more general conditions are given in Ref.~\cite{bangert}.
For such models, some exact results are known, in particular:

\begin{enumerate}
\item A ground state with a given $\ell$ exists (for some $\mu$) and is characterized, in the incommensurate case, by a strictly increasing envelope function (\ref{Ans}).
\item
It is always possible to choose $\phi$ such that
  the ground state solution $x_n$ and $n\ell + \phi$ belong to the same well of the periodic potential.
\item
  An incommensurate ground state with a given $\ell$  can  be obtained as a limit of a sequence of commensurate ground states with average bond lengths $\ell_i \rightarrow \ell$.
\item
  In any ground state with a given $\ell$, the distortions are bounded and satisfy Eq.~(\ref{inequality})~\cite{noteALD}.
\end{enumerate}

\acknowledgments
We would like to thank G. Masbaum of the Institut de Math\'ematiques de Jussieu (Paris) for stimulating discussions on various mathematical problems concerning our physical models.


\begin{thebibliography}{99}
\bibitem{Remark} Other conditions apply, in particular the perturbation must also be 'smooth enough'. If it has some singularities (possibly in higher order derivatives), KAM theorem does not apply. To know what the optimal smoothness conditions are, see Ref.~\cite{Scott}.
\bibitem{noteALD} This is proved in appendix~D of Ref.~\cite{aubry_ledaeron} for commensurate ground states (and at the limit for incommensurate ground states), formula D.2 applied for $m=n+1$.
\end{thebibliography}
\end{document}